# A dispersion-driven 3D color near-eye meta-display

Zi Wang[#,1], Dong Zhao[#,2], Li Liang[#,1], Hengyi Wang[2], Yuan Liu[1], Fang-Wen Sun[2], Kun Huang[2,3,*]

[1]National Engineering Laboratory of Special Display Technology, School of Instrumentation and Opto-Electronics Engineering, Hefei University of Technology, Hefei, Anhui 230009, China

[2]Department of Optics and Optical Engineering, University of Science and Technology of China, Hefei 230026, Anhui, China

[3]State Key Laboratory of Opto-Electronic Information Acquisition and Protection Technology, School of Physical Sciences, University of Science and Technology of China, Hefei, Anhui 230026, China

[#]These authors contributed equally to this work
[*]Corresponding authors: K. H. (huangk17@ustc.edu.cn)

**Abstract**
Chromatic dispersion, an inherent wavelength-dependent phenomenon in optical systems, has traditionally been regarded as a detrimental effect to be minimized in imaging and display. Here, we present a paradigm shift by deliberately engineering and harnessing metalens dispersion as a functional mechanism for three-dimensional (3D) near-eye displays. Specifically, we exploit lateral dispersion to transform transverse offset between green and red objects into image-space angular separations that make their images intersected virtually, thereby creating color-merged 3D virtual-image perception. This meta-display architecture preserves compactness of conventional planar display while exhibiting less data requirements and lower hardware complexity than other near-eye 3D displays. Experimentally, we demonstrate a multi-color near-eye 3D system achieving an 11° field of view, 22 pixels-per-degree angular resolution, 0.9 m depth of field, and 19 distinct image planes. This work establishes a new pathway for metasurfaces toward visual displays and highlights great potential for future virtual/augmented reality.

**Teaser**
An ultra-compact 3D multi-color near-eye display system was developed based on a metalens with lateral dispersion

**Introduction**
Optical dispersion refers to the phenomenon in which polychromatic light is separated temporally or spatially as it propagates through a material or structure, due to wavelength-dependent interactions with the medium (*1*). In most optical systems, dispersion is considered undesirable, as it generally degrades system performance. For example, in imaging systems, chromatic aberration reduces spatial resolution and blurs fine details, necessitating correction in both conventional objectives and emerging diffractive lenses (*2-7*). In optical fibre communications, material dispersion—arising from wavelength-dependent refractive indices—and waveguide modal dispersion cause temporal broadening of optical pulses (*8*). This broadening can lead to overlap between successive pulses, impairing the distinction between binary states at the receiver. Similarly, in ultrafast lasers, dispersion fundamentally limits pulse compression by introducing wavelength-dependent phase delays that broaden pulses in time (*9*). Across these applications, suppressing or compensating for dispersion—using elements such as prisms, gratings, chirped mirrors, or specialized compensating optics—is essential (*10*). Consequently, techniques for dispersion elimination have been extensively developed, underpinning advances in diverse optical technologies.

By contrast, efficient management of optical dispersion remains less developed, as scenarios demanding large dispersion are largely restricted to spectrometers (*11*) and wavelength-division multiplexing (WDM) devices (*12*) (to the best of our knowledge). In spectrometers, substantial dispersion—achieved through high diffraction orders (*13*) of gratings—is essential to spatially separate wavelengths, thus enabling high-accuracy detection and improved spectral resolution. Similarly, in WDM systems, strong dispersion helps to isolate wavelength channels spatially, thereby minimizing crosstalk between adjacent channels (*14*). Nevertheless, the full potential of dispersion remains largely untapped across broader applications. A fundamental limitation arises because the dispersion behaviour of most conventional optical elements is essentially fixed. For instance, gratings exhibit angular dispersion, diffracting different wavelengths into distinct angle positions of each non-zero order (*15*), whereas lenses with chromatic aberration typically produce longitudinally separated foci that vary strongly on wavelength (*3, 4, 16-21*). Generally, engineering such inherent dispersion in three-dimensional space freely remains challenging, which therefore considerably constrains their utility in wider applications.

To extend optical dispersion for near-eye colour displays, we present a dispersion-controlled metalens engineered to produce wavelength-dependent lateral foci by using spatial multiplexing. When imaging multiple transversely offseted objects of identical shape but distinct colours, this lateral dispersion effectively converts transverse displacements into angular separations of monochromatic light rays. This transformation enables the synthesis of a virtual, multi-colour three-dimensional image suitable for near-eye viewing. Utilizing this single metalens, our near-eye display system achieves a compact form factor, with a field of view of 11°, an angular resolution of 22 pixels per degree, an imaging depth of 0.9 m, and support for 19 distinct image planes. These performance metrics are competitive with existing three-dimensional near-eye display systems, which typically rely on more complex hardware or computationally intensive data processing. Our approach offers a route towards low-cost, ultra-compact, and high-performance 3D near-eye displays for augmented and virtual reality (*22, 23*).

**Results**
**Working principle of near-eye 3D display with a lateral-dispersion metalens**
To realize 3D color imaging, our near-eye display system (Fig. 1A) comprises three essential components: an electrically controlled display chip used to render red and green objects, a lateral-dispersion metalens designed to synthesize color 3D virtual images, and a beam splitter that guide light towards the viewer's eyes, resulting in a highly compact optical architecture. The working principle is schematically illustrated in Fig. 1B, where two pairs of dual-color objects (red and green letters "3" and "D") are illuminated by a collimated light and subsequently imaged through the metalens. The red and green objects for "3" are offset by a transverse shift $\varDelta x_1$, whereas those for the "D" are displaced by a distinct shift $\varDelta x_2$. Leveraging the lateral dispersion of the metalens, this different displacement enables the virtual synthesis of both "3" and "D" at separate longitudinal planes.

This 3D imaging capability arises primarily from the two transversely separated focal points (spaced a distance $d$ apart, Fig. 1C) of the lateral-dispersion metalens, which converges green light towards the left and red light towards the right. For any pair of corresponding red and green objects, the transverse offset $\varDelta x$ (e.g., $\varDelta x_1$ for the numeral "3") is transformed by the metalens into an angular separation $\varDelta \theta$ (e.g., $\varDelta \theta_1$) between the transmitted red and green rays, given by (see derivation in Materials and Methods)

$$\Delta \theta = \frac{d - \Delta x}{f}, \tag{1}$$

where *f* is the focal length of the metalens. This angular separation can induce 3D perception of real objects (see Note S1). After reflection by the beam splitter, the red and green rays associated with the "3" enter into human eye in a divergent manner, with their backward extensions (dashed color lines in Fig. 1B) intersecting at a virtual point *A*. Similarly, due to the different lateral shift $\Delta x_2$, the corresponding rays for the "D" converge virtually at position *B* with a distinct longitudinal distance, thereby realizing stereoscopic imaging. Mathematically, the imaging distance for each object, defined as the distance from the metalens to the virtual point (e.g., *A* and *B* in Fig. 1B) in the absence of the beam splitter (Fig. 3A), is expressed as (see derivation in Materials and Methods)

$$L = \Delta x / \Delta \theta = \frac{\Delta x \cdot f}{d - \Delta x}. \tag{2}$$

By varying $\Delta x$ from 0 to *d*, the imaging distance can be tuned from 0 to infinity, theoretically enabling a large imaging depth to reconstruct 3D scenes. In contrast, without lateral dispersion, the traditional metalens or spherical lens cannot induce 3D color imaging (see details in Note S2).

In our near-eye system, each 3D virtual image is synthesized through the combined contribution of the red and green object channels. By independently adjusting the relative intensity of each colour channel, a broad palette of hues can be generated, establishing the foundational mechanism for multi-color 3D display in our system. When these 3D color images are perceived by the human eye, the crystalline lens adjusts its focal length to project one of these virtual images sharply onto the retina, while other image planes remain perceptually blurred, thereby conferring a natural sense of depth during observation.

**Metalens with lateral dispersion**

Figure 1C illustrates the operating principle of the proposed metalens, which simultaneously focuses red ($\lambda_2$ = 660 nm) and green ($\lambda_1$ = 520 nm) light at a shared focal length of *f* = 10 mm, while introducing a lateral separation of *d* = 1 mm between the two spectral foci. This engineered lateral dispersion distinguishes our design from conventional metalenses, which typically exhibit either wavelength-dependent axial focal shifts (*24*) or achromatic behavior with spectrally overlapping foci (*3, 4, 16-21*), thereby enabling its application in near-eye 3D displays. To achieve the targeted lateral dispersion, the metalens adopts a spatially interleaved architecture comprising two metasurfaces that share the same aperture but are laterally shifted by one pixel relative to each other (Fig.1C). Each metasurface encodes a wavelength-specific phase profile given by $\varphi(x,y) = 2\pi(f - \sqrt{(x-D_\lambda)^2 + y^2 + f^2})/\lambda$, where *x* and *y* are Cartesian coordinates on the metalens plane, and the lateral offset is defined as $D_\lambda = -d/2$ at $\lambda_1 = 520$ nm and $D_\lambda = d/2$ at $\lambda_2 = 660$ nm.

To construct the metalens, nanobricks are patterned within a 300 nm-thick single-crystal silicon (c-Si) layer deposited on a sapphire substrate. For accurate phase manipulation, we employ geometric metasurfaces composed of rotating nanobricks (*25-29*), wherein the phase profiles can be precisely tailored via the in-plane rotation angle of each nanobrick. This approach exhibits superior phase accuracy compared to waveguide metasurfaces (*30*) and traditional relief-based diffractive elements (*31, 32*). Upon transmission through these nanobricks, incident circularly polarised (CP) light is convert to its orthogonal CP state, simultaneously acquiring an additional phase equal to twice the nanobrick's rotation angle (*33-35*). The interleaved architecture enables independent design of each nanobrick by using finite-difference time-domain (FDTD) method, which simulates the conversion efficiency at the designed wavelengths (see Materials and Methods for simulation details). To minimize crosstalk while preserving high optical efficiency, we numerically evaluate the conversion efficiencies of nanobricks with varying dimensions at the two operating wavelengths, $\lambda_1$ and $\lambda_2$. This analysis yields optimal nanobrick dimensions of 110 nm×80 nm×300 nm for $\lambda_1$ and 200 nm×100 nm× 300 nm for $\lambda_2$, selected to maximize the efficiency contrast between both wavelengths (see Note S3). In Fig. 1D, the simulated spectral responses of both nanobrick types exhibit a pronounced difference in conversion efficiency at each target wavelength, indicating low inter-channel crosstalk in theory and thereby enabling improved color mixing. Furthermore, to achieve sufficient imaging resolution and a wide field of view (FOV) along the focus-shifting direction (x-axis), the metalens is designed with a rectangular aperture of 1 mm×3 mm.

To experimentally validate its performance, the metalens is manufactured via standard electron beam lithography followed by a dry-etching process (see Materials and Methods for fabrication details). Figure 1E presents a microscopic image of the fabricated metalens, illustrating its rectangular outline. Fine structural details are revealed in the scanning electron microscope (SEM) image in Fig. 1F, where spatially multiplexed rotating nanobricks—corresponding to the two operating wavelengths—are highlighted in red and green. The focusing characteristics are evaluated by illuminating the metalens with collimated laser beams at red and green wavelengths (see Note S4). This produced two well-separated focal spots—red and green—with a designed lateral spacing of 1 mm at the focal plane, as shown in the recorded intensity profiles in Fig. 1G. These results confirm successful realization of the expected lateral dispersion in the metalens, without introducing significant crosstalk between red and green channels. Owing to the rectangular aperture, both focal spots exhibit an elliptical shape. To show its performance quantitatively, we compare simulated (see Materials and Methods for simulation details) and experimentally measured line-scan intensity profiles along the *x*- and *y*- directions for the two focal spots at $\lambda_1$=520 nm (Fig. 1H) and $\lambda_2$=660 nm (Fig. 1I). The close agreement between simulation and experiment verifies the design and fabrication accuracy. The measured focusing efficiencies are $\eta_1$=20.6% at $\lambda_1$=520 nm and $\eta_2$=40% at $\lambda_2$=660 nm (see measurement details in Note S5), which are slightly lower than their theoretical values $\eta_1$=27.5% and $\eta_2$=49% (see Materials and Methods) due to inevitable experimental errors from imperfect fabrication and alignment. The imaging capability of the metalens is further demonstrated in a subsequent near-eye 3D display application.

**Metalens-based near-eye 3D display**
Leveraging the good performances of our metalens, we demonstrate a proof-of-concept near-eye multi-color 3D display using a self-built optical setup illustrated in Fig. 2A. Although the system employs a laser that sequentially emits red ($\lambda_1$ = 660 nm) and green ($\lambda_2$ = 520 nm) light—chosen for its good collimation and adequate power—other low-coherence or incoherent sources such as LEDs are also compatible, as the design does not rely on coherence. Since the red-green object pairs are loaded on a reflective LCOS-SLM with a pixel pitch of 3.6 μm×3.6 μm, a 4-*f* telescopic system composed of two lenses ($L_2$ and $L_3$) is used to project the object pairs onto the image plane located at a distance *ΔS* (where *ΔS<f*) from the metalens, thereby producing magnified virtual images. We note that this 4*f* relay system can be omitted if a transmissive liquid-crystal display is placed directly at the image plane, thus making the system more compact. Owing to geometric phase used in this work, a quarter-wave plate is positioned before the metalens to generate circularly polarised incident light. Additionally, a circular polarization analyser—comprising a quarter-wave plate followed by a linear polariser—is used to suppress undesired zero-order light, see more experimental details in Materials and Methods.

Figure 2B schematically illustrates the 3D display configuration, where virtual images—an ellipsoid, a cylinder, and a cone—are superimposed onto physical panels depicting an elephant (z = 15 cm), a dog (z = 25 cm), and a lion (z = 50 cm), respectively. A camera, serving as a proxy for human vision, captures these virtual images as shown in Fig. 2C–2E. When the camera focuses on a specific longitudinal plane, the corresponding colour virtual image appears sharp with accurate spatial overlap of the red and green components. In contrast, images at other planes exhibit colour separation due to lateral chromatic offset. Notably, defocusing in this system manifests primarily as colour-dependent spatial shifts rather than conventional blurring of high-frequency details. To further verify the 3D imaging capability, we present merged virtual–real scenes in Fig. 2F–2H. As anticipated, when the camera focuses on the nearest plane (z = 15 cm), both the physical "elephant" panel and the virtual "ellipsoid" appear sharply defined (Fig. 2F), while objects at other distances—both real and virtual—are rendered out of focus. Similar results are observed when focusing on intermediate (Fig. 2G) or far planes (Fig. 2H), consistently demonstrating depth-dependent image sharpness and thereby confirming the successful reconstruction of virtual images at distinct longitudinal positions.

Multi-color creation is achieved by independently controlling the brightness of the red and green channels. When the two channels overlap with matched intensity, a sharp yellow image is perceived. By selectively enhancing the brightness of either channel, the apparent colour of the image can be shifted toward red or green. As illustrated in Fig. 2I, the red–green chromaticity coordinates measured at three points on the experimental ellipsoid pattern (Fig. 2J) confirm these perceptible colour variations. In current demonstration, the available colour space is constrained to two primary colours due to the strong absorption of single-crystal silicon in the blue spectral region. However, a three-channel metalens supporting full colour reproduction remains feasible through the adoption of alternative material platforms—such as gallium nitride, silicon nitride, or titanium dioxide—that exhibit low loss at blue wavelengths. With such materials, a multi-colour 3D display could be realized within the proposed architecture (see detailed discussions in Note S6).

**Multi-plane display**
Although the current demonstration utilizes only three image planes, the proposed methodology supports the generation of a significantly larger number of imaging planes. According to Eq. 1, the lateral object shift $\Delta x$ is converted via lateral dispersion into an angular separation $\Delta \theta$ between the red and green ray pairs, leading to the reconstructed image depth $L$ (Fig. 3A) described in Eq. 2. The imaging position can be tuned by varying $\Delta x$, which is quantized in integer multiples of the pixel pitch. The number of imaging planes depends on the lateral dispersion $d$ by following the relationship:

$$N = \left\lfloor \frac{d - \Delta x_{L_{min}}}{p} \right\rfloor, \tag{3}$$

where $\Delta x_{L_{min}}$ is the object shift corresponding to the minimum focus distance of human eye, $p$ is the pixel pitch, and the $\lfloor a \rfloor$ function means the integer part of the number $a$. Assuming a minimum distance $L$ of 12 cm and a pixel pitch of 3.6 μm, $\Delta x$ can take 21 discrete values. Thus, the display can theoretically present 21 distinct image planes, spanning a depth range from 12 cm to 356 cm (see Note S7). Notably, the depth intervals are non-uniform, widening as the depth increases. This progression aligns well with human depth perception, since visual acuity for depth discrimination decreases with distance (*36*).

For our approach, the depth of focus of metalens itself almost has no influence on the imaging depth. Each monochromatic viewpoint is implemented via a retinal projection display scheme (*37-39*), which provides an always-in-focus characteristic (Note S8). In this configuration, a collimated backlight produces narrow optical rays for each pixel; these rays are directly converged and projected onto the retina. As such narrow rays are largely insensitive to focal changes of the eye lens, the effective imaging depth in this system can be considered sufficiently large.

Figure 3B illustrates the principle for evaluating image depth experimentally. Two printed patterns—an "elephant" and a "dog"—serve as reference objects. The elephant remains fixed at a distance of 15 cm, while the dog is longitudinally translated to align with the depth of the virtual ellipsoid image. In Fig. 3C, five augmented-reality scenes are captured with the camera focused at distances of 15 cm, 23 cm, 31 cm, 74 cm, and 104 cm, respectively. A progressive reduction in the apparent size of the dog from left to right indicates increasing imaging distance. At each focal setting, both the dog and the virtual ellipsoid remain in sharp focus, confirming their depth correspondence. The position of the virtual image is thus determined indirectly by the measured position of the dog. Figure 3D plots the theoretical and experimentally measured relationship between the object shift $\Delta x$ and the imaging distance $L$. The close agreement between data and theory confirms the capability of the display to render multiple image planes across a broad depth. This is further demonstrated in Fig. 3E, which shows eight animal patterns reconstructed at distinct distances. When the camera focuses at a specific plane, only the corresponding pattern appears sharp, while others exhibit noticeable defocus blur.

To evaluate the performance limits of the system, we experimentally reconstruct 19 images featuring a "square" pattern at different longitudinal planes, spanning depths from 12 cm to 99 cm (Note S9). Although the system theoretically supports the reconstruction of 21 images, the large longitudinal interval between the 20$^{th}$ and 21$^{st}$ planes prevented their experimental observation, leaving these two images undemonstrated. As shown in Supplementary fig. S9A, when the camera is focused on the first plane, the 19 squares exhibit distinct blur sizes, confirming their reconstruction at different depths.

To evaluate the dynamic capabilities of the system, we experimentally reconstruct a rotating ball moving along a three-dimensional trajectory—from a distant starting point toward the camera's focal plane. Selected frames from a video capturing this 3D motion are presented in Fig. 3F, with the full sequence available in Supplementary Movie S3. When the camera lens is focused at a short distance, the object exhibits increasing sharpness as it approaches the camera, demonstrating the capacity of our method to represent motion in 3D space. To further verify it, an additional animation provided in Supplementary Movie S4 shows the rotating ball captured with the camera lens focused at a far distance.

**Resolution and FOV of our near-eye display**

To evaluate resolution across different depths, we reconstruct concentric square patterns at varying distances, as depicted in Fig. 4A. The image size increases with reconstruction depth, a result of the nearly constant field of view (FOV). Physical dimensions are measured using a ruler. From each image, we extract the intensity profile along the horizontal line crossing the pattern center and computed its Fourier amplitude spectrum (Fig. 4B). For the initial spectrum, the principal frequency is identified as 10.53 cycles per degree, corresponding to an angular resolution of 21.06 pixels per degree (see Materials and Methods for calculation details). Angular resolutions at other depths are determined following the same procedure. As summarized in Fig. 4C, the angular resolution remains relatively consistent across a wide depth range, averaging 21.86 pixels per degree. This stability supports high-quality three-dimensional reconstruction over varied imaging depths.

Figure 4D illustrates the approach used to evaluate the field of view (FOV). The green and red viewpoints represent the rightward and leftward FOVs, respectively, with their overlapping yellow region defining the effective FOV of the display system. In principle, its maximum FOV is calculated as:

$$FOV = 2\tan^{-1}\left[\frac{D(L+f) - dL}{2f(L+f)}\right] \qquad (4)$$

where $D$ is the aperture of the metalens. A smaller lateral dispersion $d$ leads to a larger FOV but reduces the number of image planes, revealing a fundamental trade-off between these two parameters. When a metalens with a large aperture $D$ is employed, the influence of lateral dispersion $d$ on the FOV becomes less pronounced, allowing dispersion to be strategically increased to achieve more image planes. In Fig. 4E, a yellow grid pattern reconstructed at an imaging distance $L$=12 cm measures 25 mm in length, corresponding to an experimental FOV of $2\tan^{-1}\frac{2.5}{2\times(12+1)} \approx 11°$. Figure 4F shows that the theoretical FOV remains nearly constant, with minor variations around 11.4°. Experimentally measured FOV values are slightly lower than theoretical predictions, which is caused by unavoidable experimental uncertainties.

**Discussion**
After decades of development, mainstream 3D near-eye display systems have converged into four principal categories: holography (*23, 40-47*), multi-plane displays (*48-51*), light field displays (*52, 53*) and super multi-view systems(*39*). Although each approach offers distinct advantages, they collectively face several shared challenges: high hardware requirements, substantial computational and data throughput demands, and considerable system complexity and bulk. A comparative analysis of hardware architecture characteristics between the proposed method and mainstream 3D near-eye displays is summarized in Table 1. As indicated, holographic, multi-plane, and super multi-view systems are characterized by high architectural complexity and large physical volume. Moreover, holography and super multi-view approaches exhibit limited compatibility with conventional light sources. Although light field displays achieve a compact form factor and broad source compatibility, they typically necessitate high-resolution panels owing to the inherent spatial–angular resolution trade-off. These methods also incur substantial computational and bandwidth overhead from processing, storing, and transmitting wave-optical or light field data. In contrast, the proposed system attains minimal complexity, reduced computational and bandwidth requirements, and a compact footprint, while retaining high compatibility with standard light sources.

Despite these advantages, several aspects of our current system require further optimization. Replacing the current LCOS-SLM with a transmissive LCD would substantially reduce the system volume, achieving the compact form factor illustrated in Fig. 1A. Furthermore, employing a well-fabricated, large-aperture metalens is expected to enhance key performance metrics—including FOV, imaging depth, and the number of image planes—to levels competitive with other 3D near-eye displays (see Note S10). Image quality could also be improved by adopting an LED source to suppress laser speckle noise. Additionally, the current system exhibits a restricted eyebox of 3 mm (roughly estimated by $d_p$-$d$, where $d_p$ is pupil diameter, and taken as 4 mm on average). A potential solution involves the design of a metalens capable of generating multiple discrete red and green focal points, as conceptualized in Supplementary fig. S10.

In summary, we have developed a 3D multi-color near-eye display system based on a metalens with lateral dispersion. Rather than suppressing chromatic dispersion—typically regarded as a drawback—we strategically engineered it to angularly separate distinct wavelengths, thereby generating monocular accommodation cues on the retina. The design avoids complex optical components or specialized hardware, ensuring a compact form factor and low implementation barrier. Moreover, the minimal data requirements enable real-time capture, processing, transmission, and reconstruction of 3D content. This dispersion-driven approach might offer a practical and efficient pathway toward future 3D near-eye imaging and display (*41, 54-56*).

**Materials and Methods**
**Simulation details about calculating conversion efficiency of nanobricks**
A finite-difference time-domain (FDTD) method is employed to evaluate the conversion efficiency of silicon nanobricks. With target wavelengths set at 520 nm and 660 nm, a unit cell period of 250 nm × 250 nm—approximately half the wavelength—is selected to enhance conversion efficiency. The nanobrick height is fixed at 300 nm, corresponding to the thickness of the crystalline silicon (c-Si) film used in our sapphire-supported platform. Periodic boundary conditions are applied along the x- and y-directions, while a perfectly matched layer (PML) is implemented along the z-axis. By systematically varying the in-plane dimensions of the nanobricks, the corresponding conversion efficiencies at 520 nm and 660 nm are numerically determined. Supplementary fig. S3 shows the simulated conversion efficiencies for size-different nanobricks at the designed wavelengths of 520 nm and 660 nm. Considering the interleaved architecture in our fabricated metalens, its theoretical efficiencies are half of the simulated values at both designed wavelengths, i.e., the theoretically efficiencies of our interleaved metalens are $\eta_1$=55%/2=27.5% at $\lambda_1$= 520 nm and $\eta_2$=98%/2=49% at $\lambda_2$= 660 nm.

## Numerical simulation about optical focusing with metalens

The metalens features a rectangular aperture of 1 mm × 3 mm in the metasurface plane. Its phase profile is defined as $\varphi(x,y)= 2\pi(f-\sqrt{(x-D_\lambda)^2+y^2+f^2})/\lambda$, as provided in the main text, while the amplitude is derived from FDTD simulations using specifically selected nanobricks. The interleaved complex amplitude distributions at different wavelengths are used to separately compute the corresponding focal spots in the focal plane. Light propagation is modelled via the Rayleigh–Sommerfeld integral, implemented using a fast Fourier transform algorithm (*24*). Finally, simulated line-scan intensity profiles (Figs. 1H, 1I) are extracted along the centre of the focal spots, with two-dimensional representations provided in Supplementary fig. S4B.

## Fabrication of the metalens

The metalens was fabricated using electron-beam lithography followed by dry etching. A positive-tone electron-beam resist (AR-P 6200) was spin-coated and baked onto a 300 nm-thick silicon film grown on a sapphire substrate. Patterning was performed via electron-beam lithography (JEOL JBX 6300FS) operating at 100 kV. After development, the sample was post-baked to remove residual moisture. A 10 nm-thick chromium hard mask was then deposited using electron-beam evaporation (Kurt J. Lesker PVD75 Proline). Following lift-off, the silicon layer was etched using an inductively coupled plasma etching system (Oxford PlasmaPro System100 ICP380). Finally, the remaining chromium mask was stripped with a chromium etchant.

## Experimental setup of the 3D meta-display

To experimentally validate the proposed approach, we constructed a near-eye 3D display prototype incorporating the metalens design illustrated in Fig. 2A. A collimated laser source illuminates a liquid crystal on silicon spatial light modulator (LCOS-SLM; UPOLabs HDSLM36RA), with a pixel size of 3.6 μm, resolution of 3840 × 2160, and a refresh rate of 180 Hz. Owing to the absence of a colour filter on the LCOS-SLM, red and green laser illumination is sequentially switched, while corresponding image patterns are synchronously loaded onto the SLM. A 4*f* relay system projects the image plane within the focal range of the metalens, enabling the formation of magnified virtual images. The metalens with a size of 3 mm × 1 mm has a focal length of 10 mm. A quarter-wave plate converts the incident linearly polarised light into left-circularly polarised light. Upon interaction with the metalens, part of this left-circularly polarised component is converted into a right-circularly polarised state and is focused to form two laterally separated red and green focal points spaced 1 mm apart. The remaining left-circularly polarised light transmits through the metalens and is subsequently blocked by a right-circular analyser.

## Transformation between object shift (Δx) and angular separation (Δθ)

As illustrated in Fig. 3A, the angular separation $\Delta\theta$ is sufficiently small to permit the following approximations:

$$\Delta\theta = \frac{\Delta x}{L} \tag{5}$$

$$\Delta\theta = \frac{d}{L+f} \tag{6}$$

Eliminating the variable $L$ from equations (5) and (6) yields the relationship between the object shift $\Delta x$ and the angular separation $\Delta\theta$:

$$\Delta\theta = \frac{d - \Delta x}{f} \quad (7)$$

Extending the analysis to two dimensions, the coordinates of the two object points required to reconstruct an image point at $(x_0, L)$ are given by:

$$x_1 = \frac{f}{L+f}\left(x_0 + \frac{d}{2}\right) - \frac{d}{2} \quad (8)$$

$$x_2 = \frac{f}{L+f}\left(x_0 - \frac{d}{2}\right) + \frac{d}{2} \quad (9)$$

**Calculation of angular resolution**

To quantitatively assess resolution at various depths, we implemented a systematic procedure involving pixel intensity analysis and Fourier-domain characterization. For each reconstructed concentric-square pattern, a one-dimensional intensity profile $s(x)$ was extracted along the $x$-axis. To minimize artifacts in subsequent frequency analysis, $s(x)$ was pre-processed to generate a windowed signal $s_w(x)$ through the following steps.

First, the DC component was removed by subtracting the mean intensity $s_1$, defined as:

$$s_1 = \frac{1}{N_x} \sum_{x=0}^{N_x-1} s(x) \quad (10)$$

where $N_x$ denotes the total number of pixels in the profile. A Hanning window $w(x)$ is then applied:

$$w(x) = 0.5\left[1 - \cos\left(\frac{2\pi x}{N_x - 1}\right)\right], x = 0, 1, \ldots, N_x - 1 \quad (11)$$

yielding the pre-processed signal:

$$s_w(x) = [s(x) - s_1] * w(x) \quad (12)$$

The windowed signal $s_w(x)$ undergoes a discrete Fourier transform to obtain its frequency distribution $S(f)$, where $f$ represents the frequency index ($f = 0, 1, \ldots, N_x/2$).

$$S(f) = \sum_{x=0}^{N_x-1} s_w(x) e^{-\frac{j2\pi f x}{N_x}}, \ |S(f)| \ for \ f = 0, 1, \ldots, (N_x/2) \quad (13)$$

We computed the amplitude spectrum $|S(f)|$ for non-negative frequencies $f=0,1,\ldots, N_x/2$ to identify dominant spatial frequencies. To convert the discrete frequency $f$ (in cycles per pixel) into angular frequency $f_d$ (in cycles per degree), we applied:

$$f_d = \left(\frac{f}{W}\right) * d * \frac{\pi}{180} \quad (14)$$

where $W$ is the physical width of the image along the x-axis (in mm), and $d$ is the imaging depth in units of millimeter. This angular frequency spectrum enables consistent resolution evaluation across different depths and spatial scales

**References**
1. B. E. A. Saleh, & Teich, M. C., Fundamentals of Photonics (3rd ed.). *John Wiley & Sons.*, (2019).
2. M. Khorasaninejad, F. Capasso, Metalenses: Versatile multifunctional photonic components. *Science* **358**, (2017).

**Acknowledgements**
We thank the support from the University of Science and Technology of China's Centre for Micro and Nanoscale Research and Fabrication. The numerical calculations were partially performed on the supercomputing system at Hefei Advanced Computing Center and the Supercomputing Center of the University of Science and Technology of China.

**Funding**
National Key Research and Development Program of China: 2022YFB3607300 (K. H), 2023YFB2806800 (Z. W);
National Natural Science Foundation of China: Grant Nos. 62275071 (Z. W), 62322512, 62225506, 62505308 and 12134013;
Fundamental Research Funds for the Central Universities: WK2030000108, WK2030000090;
CAS Project for Young Scientists in Basic Research: Grant No.YSBR-049);
China Postdoctoral Science Foundation: 2023M743364 (D. Z)
Anhui Natural Science Foundation: 2508085QA010 (D. Z).

**Author Contributions**
K. H. and Z. W. conceived the idea. D. Z., F. S. and K. H. designed the metalens. D. Z. and H. W. prepared and fabricated the samples. Z. W., L. L. and Y. L. built the experimental setup and implemented the measurement. Z. W., D. Z. and K. H. analyzed the data and wrote the manuscript. K. H. supervised the overall project. All the authors discussed the results.

**Competing interests**
The authors declare that they have no competing interests.

**Data and materials availability:**
All data needed to evaluate the conclusions in the paper are present in the paper and/or the Supplementary Materials

**Figures and Tables**

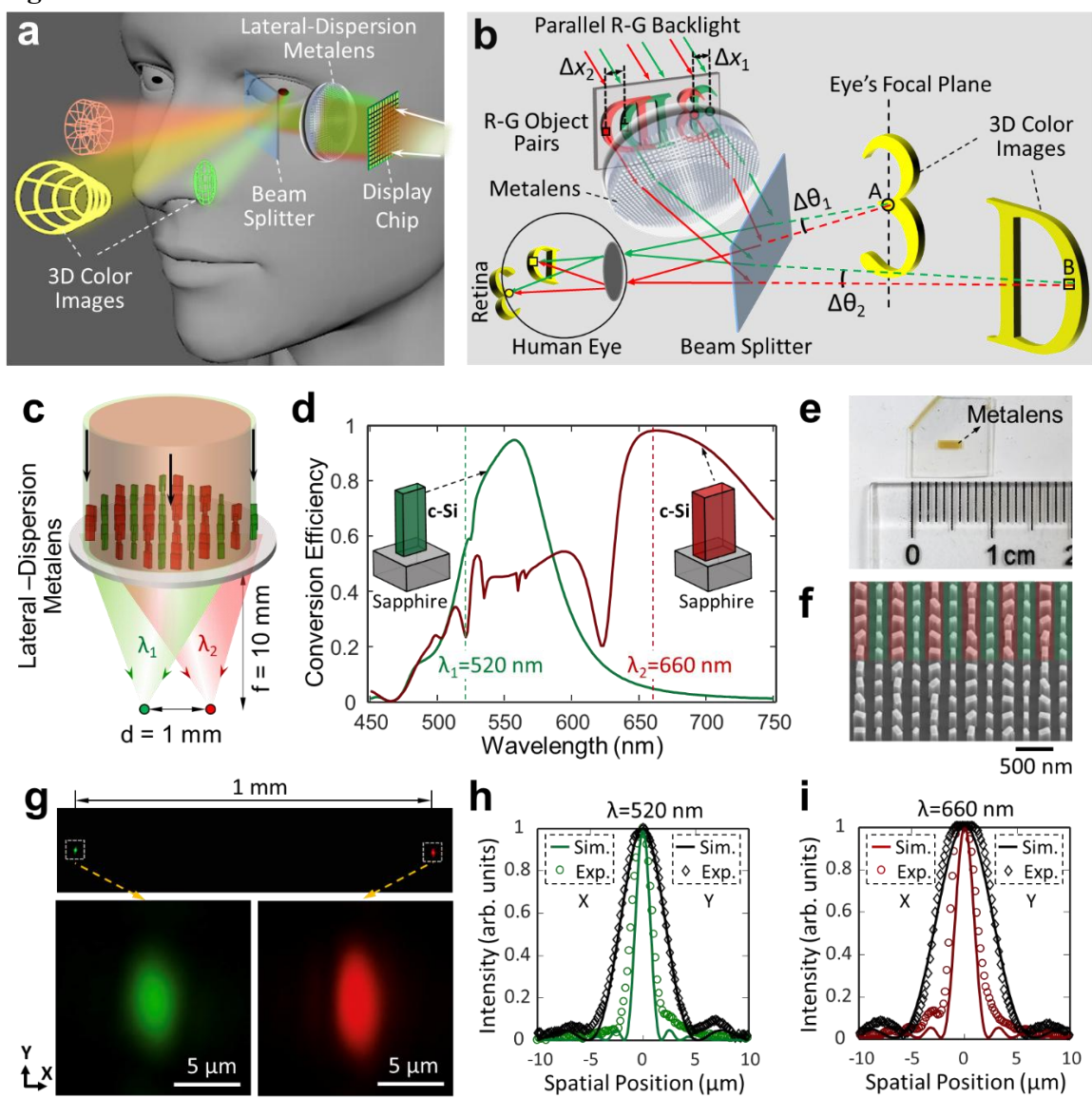

**Fig. 1. Working principle for near-eye 3D color meta-display by using a lateral-dispersion metalens. (A)** Architecture of our metalens-based 3D near-eye display. Only a metalens and a conventional parallel-lit display are used to reconstruct 3D images. **(B)** Illustration of the procedure for the reconstruction of a 3D image. The metalens uses lateral dispersion to transform object's transverse shift (*Δx*) into angular separation (*Δθ*) of transmitted green and red rays at the pupil, thereby reconstructing image points at corresponding depths. **(C)** Sketch for lateral dispersion of the metalens focusing green and red light into two transversely separated foci at the focal plane. *f*=10 mm and *d*=1 mm are used in this work. **(D)** Simulated conversion efficiencies of two silicon nanobricks in a wide spectrum from 450 nm to 750 nm. The nanobricks has the dimensions of 110 nm×80 nm×300 nm at $\lambda_1$ and 200 nm×100 nm× 300 nm at $\lambda_2$. **(E)** Optical images of the fabricated metalens with a rectangular size of 1 mm×3 mm. **(F)** Top-view SEM image of nanobricks in the metalens. Nanobricks operating at different wavelengths $\lambda_1$ and $\lambda_2$ are labelled in green and red. **(G)** Measured intensity profiles of two separated red and green focal spots via illuminating the metalens with a collimated light. The zoomed-in images are shown at the bottom panel for a better observation. **(H-I)** Normalized intensity profiles along the spots center in experiment and simulation for two wavelengths $\lambda_1$ **(H)** and $\lambda_2$ **(I)**.

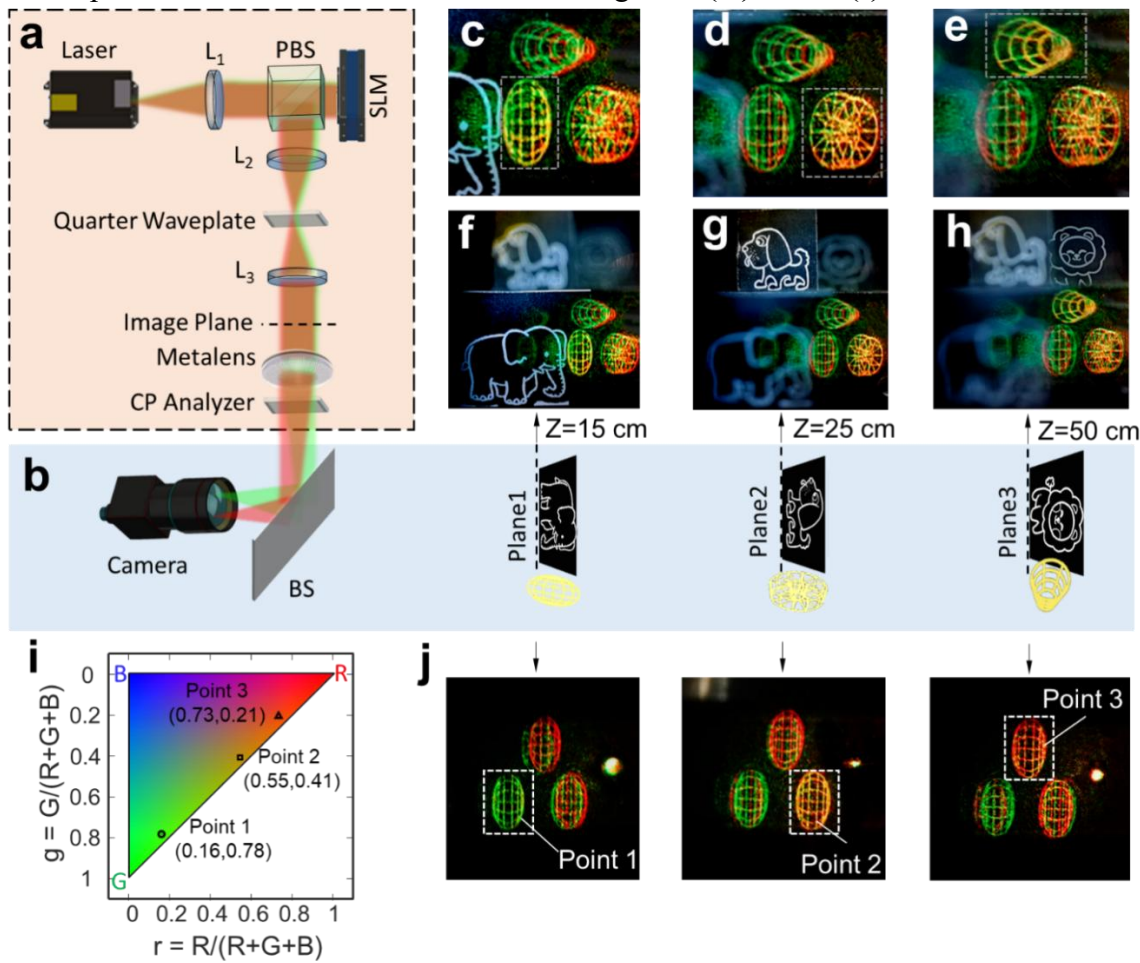

**Fig. 2. Color near-eye 3D augmented-reality display. (A)** Sketch for experimental setup of our color near-eye 3D display using the proposed metalens. L: Lens; PBS: Polarization beam splitter; CP: circular polarization; SLM: spatial light modulator. **(B)** Diagram for capturing the 3D augmented-reality images consisting of both real and virtual images. **(C-E)** Enlarged details of virtual images at three imaging planes of z=15 cm **(C)**, z=25 cm **(D)** and z=50 cm **(E)**. **(F-H)** Experimentally captured augmented-reality images at z=15 cm **(F)**, z=25 cm **(G)** and z=50 cm **(H)**. Supplementary Movie S1 presents the recorded video demonstrating its dynamic display. **(I)** The red-green chromaticity coordinates of three points demonstrated on the ellipsoid patterns in **(J)**. **(J)** Three measured "ellipsoids" with different colors (green, yellow and red) that are reconstructed at three different imaging planes.

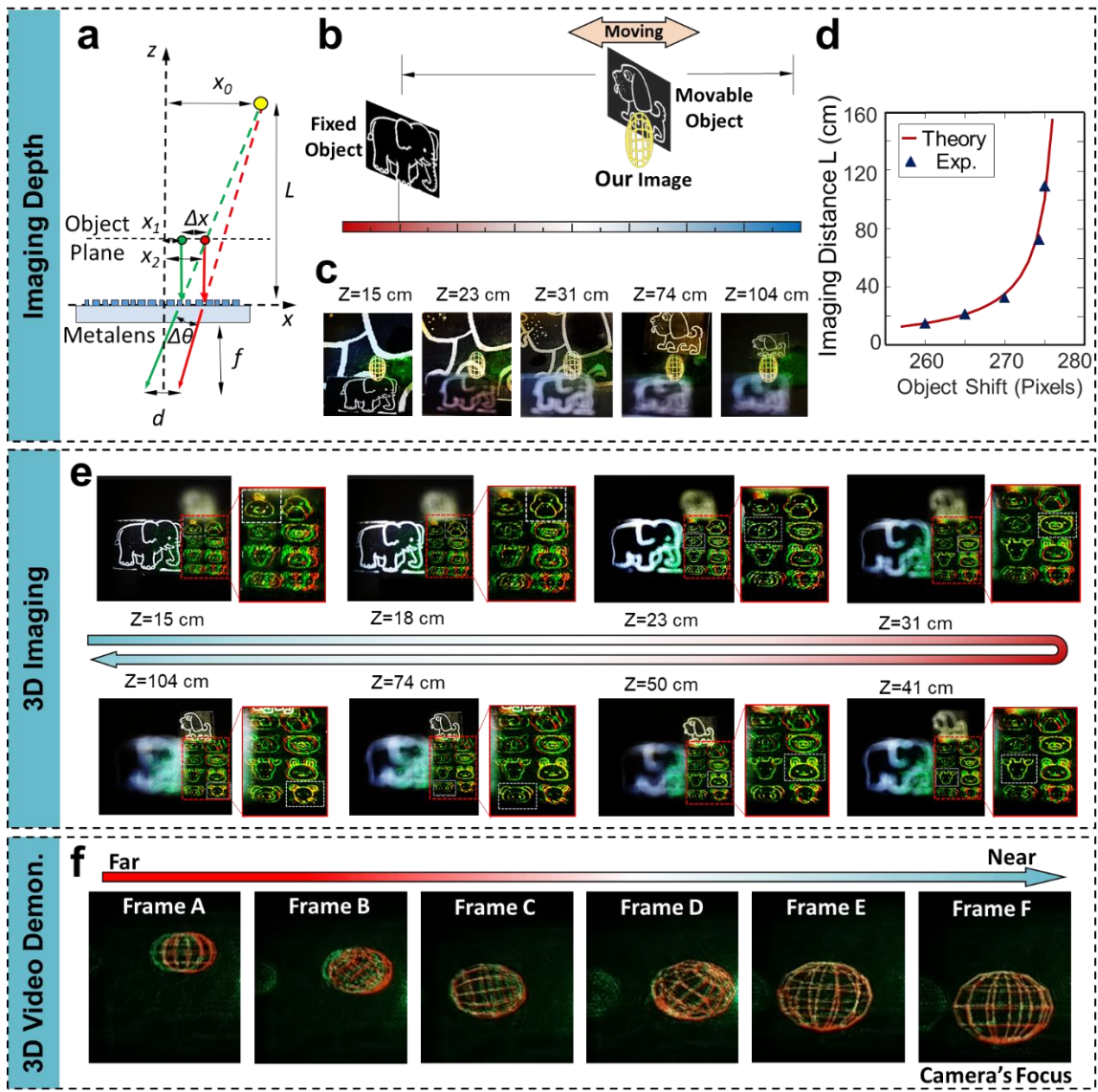

**Fig. 3. Multi-plane color display. (A)** Illustration of relationship among imaging distance $L$, object shift $\Delta x$ and angle separation $\Delta \theta$ of the red- and green-ray pairs. The green and red points at the object plane are located at $x=x_1$ and $x=x_2$, respectively. **(B)** Sketch for experimental implementation of measuring 3D depth of field. The movable object matches the depth of the virtual image "ellipsoid". **(C)** Measured virtual images of "ellipsoid" at different imaging distances when the real "elephant" is fixed at $z=15$ cm and the real "dog" pattern with fixed dimension moves from $z=15$ cm to $z=104$ cm. As $z$ increases, the blurring "elephant" and the shrunken "dog" indicate experimental acquisition of 3D virtual imaging at different longitudinal distances. **(D)** Theoretical (red solid curve) and experimental (triangles) imaging distance $L$ for different object shift $\Delta x$. **(E)** Experimental demonstration of a 3D color display at different imaging planes from $z=15$ cm to $z=104$ cm. Supplementary Movie S2 records its dynamic display when the camera's focus is tuned. **(F)** Selected frames of a 3D video demonstrating a rotating sphere that moves along a curved trajectory from a distant position to the camera's focusing point.

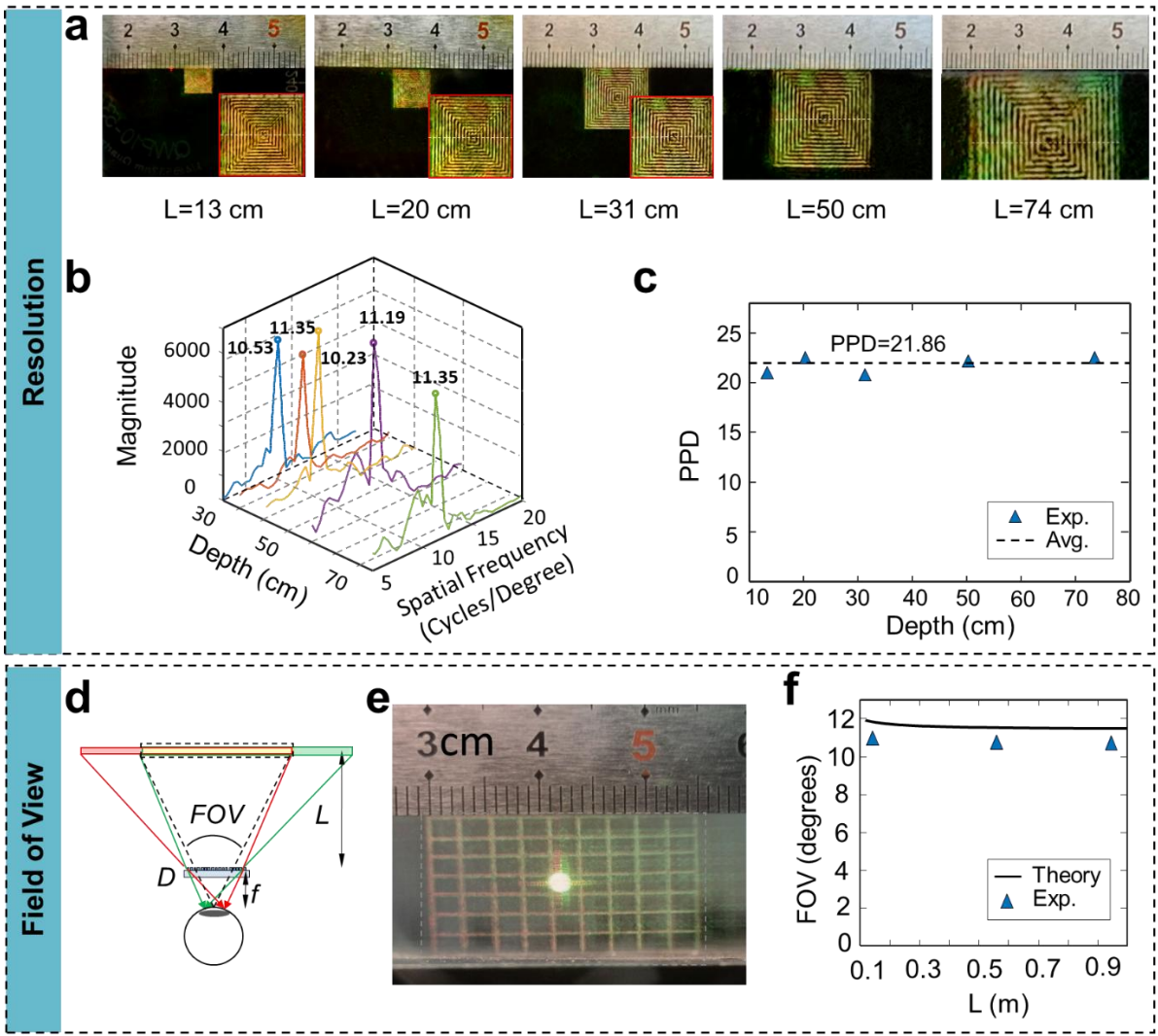

**Fig. 4. Imaging resolution and FOV of our demonstrated near-eye 3D display. (A)** 3D resolution test by projecting concentric squares at exemplified depths from $L$=13 cm to $L$=74 cm. The zoomed-in patterns are provided in the insets for the cases at $L$=13 cm, $L$=20 cm and $L$=31 cm. **(B)** Fourier amplitude spectra of the intensity distributions along the white dotted lines for all cases in **(A)**. **(C)** Measured angular resolution (in units of pixels per degree) obtained from the Fourier spectra at different depths. **(D)** Sketch for measuring actual FOV of our near-eye 3D display system. **(E)** Measured "grid" pattern for FOV characterization at an imaging distance of $L$=12 cm. The central bright spot comes from the unconverted background light that can be suppressed by designing higher-efficiency metasurfaces. **(F)** Theoretical (solid curve) and experimental (triangles) FOVs at different imaging distances.

**Table 1 Hardware comparison between our method and mainstream 3D near-eye approaches**

| Approaches | Display Hardware Barriers | System Complexity | Data Resource Consumption (computation/ | Volume | Light Source Requirement |
|---|---|---|---|---|---|

| | | | storage/bandwidth) | | |
|---|---|---|---|---|---|
| Holography | High (Phase/ Complex modulation) | High (4-f system) | High | Large | High (Lasers) |
| Multi-Plane Dsplay | Medium (High refresh-rate) | High (Focus-tunable element and synchronization) | Medium | Large | Low (Lasers or Incoherent Sources) |
| Light Field | Medium (High resolution) | Low | Medium | Small | Low (Lasers or Incoherent Sources) |
| Super Multi-view | Medium (High refresh-rate) | High (Light steering element and synchronization) | Medium | Large | High (Lasers) |
| Our Method | Low | Low | Low | Small | Low (Lasers or Incoherent Sources) |